\documentclass[aps,twocolumn,groupedaddress,nofootinbib]{revtex4}
\usepackage{graphicx}
\usepackage{amsmath}
\usepackage{epsfig}

{
{
\begin{document}

\title{Leptogenesis, Dark Matter and Higgs Phenomenology at TeV}

\author{Pei-Hong Gu$^{1}_{}$}

\author{Utpal Sarkar$^{2}_{}$}

\affiliation{$^{1}_{}$The Abdus Salam International Centre for
Theoretical Physics, Strada Costiera 11, 34014 Trieste, Italy\\
$^{2}_{}$Physical Research Laboratory, Ahmedabad 380009, India}

\begin{abstract}
We propose an interesting model of neutrino masses to realize
leptogenesis and dark matter at the TeV scale. A real scalar is
introduced to naturally realize the Majorana masses of the
right-handed neutrinos. We also include a new Higgs doublet that
contributes to the dark matter of the universe. The neutrino masses
come from the vacuum expectation value of the triplet Higgs scalar.
The right-handed neutrinos are not constrained by the neutrino
masses and hence they could generate leptogenesis at the TeV scale
without subscribing to resonant leptogenesis. In our model, all new
particles could be observable at the forthcoming Large Hardon
Collider or the proposed future International Linear Collider.

\end{abstract}

\maketitle

\textit{Introduction}: Currently many neutrino oscillation
experiments \cite{pdg2006} have confirmed that neutrinos have tiny
but nonzero masses. This phenomenon is elegantly explained by the
seesaw mechanism \cite{minkowski1977}, in which neutrinos naturally
acquire small Majorana \cite{minkowski1977} or Dirac \cite{mp2002}
masses. At the same time, the observed matter-antimatter asymmetry
\cite{pdg2006} in the universe can be generated via leptogenesis
\cite{mp2002,fy1986,luty1992,fps1995,pilaftsis1997,ms1998,ars1998,di2002,guo2004,bpy2005}
in these models. In this scheme, the right-handed neutrinos or other
virtual particles need have very large masses to realize a
successful leptogenesis if we don't resort to the resonant effect
\cite{fps1995,pilaftsis1997} by highly fine tuning.

Another big challenge to the standard model is the dark matter
\cite{pdg2006}. What is the nature of dark matter? Recently, it has
been pointed out \cite{ma06,bhr2006,ma2006,co2006} that a new Higgs
doublet can be a candidate for the dark matter if it doesn't decay
into the standard model particles. Although the possibility of Higgs
doublet to be a dark matter candidate was proposed many years back
\cite{ma77}, following the recent proposal \cite{ma06} a thorough
analysis have been carried out \cite{bhr2006,hnot2006} demonstrating
its consistency with all the recent results. In this interesting
scenario, the dark matter is expected to produce observable signals
at the Large Hardon Collider (LHC) \cite{bhr2006} and in the GLAST
satellite experiment \cite{hnot2006}. Combining this idea, the
type-I seesaw and the concept \cite{kuzmin1997} of generation of the
cosmological matter-antimatter asymmetry along with the cold dark
matter, the author of \cite{ma2006} successfully unified the
leptogenesis and dark matter. However, this scenario need the
right-handed neutrinos to be very heavy, around the order of
$10^{7}_{}\, \textrm{GeV}$.

In this paper, we propose a new scheme to explain neutrino masses,
baryon asymmetry and dark matter at TeV scale by introducing a Higgs
triplet which is responsible for the origin of neutrino masses, a
new Higgs doublet that can be a candidate for the
dark matter, and a real scalar which can generate the
Majorana masses of the right-handed neutrinos naturally.
A discrete symmetry ensures that the new Higgs doublet cannot
couple to ordinary particles. This same discrete symmetry
will also prevent any connection between the right-handed neutrinos
and left-handed neutrino masses. This allows the right-handed
neutrinos to decay at low scale generating the lepton
asymmetry, which will be finally converted to the baryon asymmetry
through the sphaleron processes \cite{krs1985}. This will then explain the
observed matter-antimatter asymmetry in the universe, even if the
Majorana masses of the right-handed neutrinos
are not highly quasi-degenerate. In our model, all
new particles could be close to the TeV scale and hence should be
observable at the forthcoming LHC or the proposed future
International Linear Collider (ILC).

\textit{The model}: We extend the standard model with some new
fields. The field content is shown in Table \ref{charge}, in which
\begin{eqnarray}
\psi_{L}^{} &=& \left \lgroup \begin{array}{c}
\nu_{L}^{}\\
l_{L}^{}\end{array} \right \rgroup\,,\quad \phi = \left \lgroup
\begin{array}{c}
\phi^{0}_{}\\
\phi^{-}_{}\end{array} \right \rgroup
\end{eqnarray}
are the left-handed lepton doublet and Higgs doublet of the standard
model, respectively, while
\begin{eqnarray}
\eta = \left \lgroup
\begin{array}{c}
\eta^{0}_{}\\
\eta^{-}_{}\end{array} \right \rgroup
\end{eqnarray}
is the new Higgs doublet that will be the dark matter candidate,
$\nu_{R}^{}$ is the right-handed neutrino, $\chi$ is the real scalar
and
\begin{eqnarray}
\Delta_{L}^{} &=& \left \lgroup
\begin{array}{cc}
\frac{1}{\sqrt{2}}\delta^{+}_{} &\delta^{++}_{}\\
\delta^{0}_{} &-\frac{1}{\sqrt{2}}\delta^{+}_{}\end{array} \right
\rgroup
\end{eqnarray}
is the Higgs triplet. We further introduce a discrete $Z_{4}^{}$
symmetry, under which the different fields transform as
\begin{eqnarray}
&&\psi_{L}^{}\,\,\rightarrow \,\psi_{L}^{}\,,\quad \phi
\,\,\rightarrow
\,\,\,\,\,\phi\,,\quad\eta\,\,\,\,\,\,\,\rightarrow -i\eta\,,\nonumber\\
&&\nu_{R}^{}\,\,\rightarrow i\nu_{R}^{}\,,\quad\chi\,\,\rightarrow
-\chi\,,\quad\Delta_{L}^{}\,\,\rightarrow\,\,\,\,\,\,\Delta_{L}^{}\,.
\end{eqnarray}
Here the other standard model fields, which are all even under the
$Z_{4}^{}$, and the family indices have been omitted for simplicity.

\begin{table}
\begin{center}
\begin{tabular}{c|cccccc}
\hline\hline
\\[-2.5mm]
\quad Fields \quad \quad       & \quad ~\,$\psi_{L}^{}$ \quad &
\quad ~\,$\phi$ \quad & \quad ~\,$\eta$\quad & \quad
~\,$\nu_{R}^{}$\quad & \quad ~\,$\chi$\quad & \quad
~\,$\Delta_{L}^{}$\quad\quad
\\[2.5mm]
\hline
\\[-2.5mm]
$SU(2)_{L}^{}$  & \quad   ~\,\textbf{2}    \quad &
\quad~\,\textbf{2}\quad & \quad~\,\textbf{2}\quad
&\quad~\,\textbf{1}\quad &\quad~\,\textbf{1}\quad
&\quad~\,\textbf{3}\quad\quad
\\[2.0mm]
$U(1)_{Y}^{}$         &   \quad  $ -\frac{1}{2} $    \quad & \quad$
-\frac{1}{2} $\quad & \quad$ -\frac{1}{2} $\quad&\quad~\, $ 0
$\quad& \quad~\,$ 0 $\quad& \quad~\,$ 1 $\quad\quad
\\[-2.5mm]
\\ \hline \hline
\end{tabular}
\caption{The field content in the model. Here $\psi_{L}^{}$, $\phi$ are the
standard model left-handed lepton doublets and Higgs doublet, $\eta$
is the new Higgs doublet, $\nu_{R}^{}$ is the right-handed
neutrinos, $\chi$ is the real scalar and $\Delta_{L}^{}$ is the
Higgs triplet. Here the other standard model fields and the family
indices have been omitted for simplicity.} \label{charge}
\end{center}
\end{table}

We write down the relevant Lagrangian for the Yukawa interactions,
\begin{eqnarray}
\label{lagrangian1} -\mathcal{L}&\supset&
\sum_{ij}^{}\left(y_{ij}^{}\overline{\psi_{Li}^{}}\eta
\nu_{Rj}^{}+\frac{1}{2}g_{ij}^{}\chi\overline{\nu_{Ri}^{c}}\nu_{Rj}^{}\right.\nonumber\\
&+&\left.\frac{1}{2}f_{ij}^{}\overline{\psi_{Li}^{c}}i\tau_{2}^{}\Delta_{L}^{}\psi_{Lj}^{}+\textrm{h.c.}\right)\,,
\end{eqnarray}
where $y_{ij}^{}$, $g_{ij}^{}$, $f_{ij}^{}$ are all dimensionless.
We also display the general scalar potential of $\phi$, $\eta$,
$\chi$ and $\Delta_{L}^{}$,
\begin{eqnarray}
\label{potential}
&&V(\chi,\phi,\eta,\Delta_{L}^{})\nonumber\\
&=&\frac{1}{2}\mu_{1}^{2}\chi^{2}_{}+\frac{1}{4}\lambda_{1}^{}\chi^{4}_{}+\mu_{2}^{2}\left(\phi^{\dagger}_{}\phi\right)+\lambda_{2}^{}(\phi^{\dagger}_{}\phi)^{2}_{}\nonumber\\
&+&\mu_{3}^{2}\left(\eta^{\dagger}_{}\eta\right)+\lambda_{3}^{}(\eta^{\dagger}_{}\eta)^{2}_{}+M^{2}_{\Delta}\textrm{Tr}\left(\Delta_{L}^{\dagger}\Delta_{L}^{}\right)\nonumber\\
&+&\lambda_{4}^{}\textrm{Tr}\left[\left(\Delta_{L}^{\dagger}\Delta_{L}^{}\right)^{2}_{}\right]+\lambda_{5}^{}\left[\textrm{Tr}\left(\Delta_{L}^{\dagger}\Delta_{L}^{}\right)\right]^{2}_{}\nonumber\\
&+&\alpha_{1}^{}\chi^{2}_{}\left(\phi^{\dagger}_{}\phi\right)+\alpha_{2}^{}\chi^{2}_{}\left(\eta^{\dagger}_{}\eta\right)+\alpha_{3}^{}\chi^{2}_{}\textrm{Tr}\left(\Delta_{L}^{\dagger}\Delta_{L}^{}\right)\nonumber\\
&+&2\beta_{1}^{}\left(\phi^{\dagger}_{}\phi\right)\left(\eta^{\dagger}_{}\eta\right)+2\beta_{2}^{}\left(\phi^{\dagger}_{}\eta\right)\left(\eta^{\dagger}_{}\phi\right)\nonumber\\
&+&2\beta_{3}^{}\left(\phi^{\dagger}_{}\phi\right)\textrm{Tr}\left(\Delta_{L}^{\dagger}\Delta_{L}^{}\right)+2\beta_{4}^{}\phi^{\dagger}_{}\Delta_{L}^{\dagger}\Delta_{L}^{}\phi\nonumber\\
&+&2\beta_{5}^{}\left(\eta^{\dagger}_{}\eta\right)\textrm{Tr}\left(\Delta_{L}^{\dagger}\Delta_{L}^{}\right)+2\beta_{6}^{}\eta^{\dagger}_{}\Delta_{L}^{\dagger}\Delta_{L}^{}\eta\nonumber\\
&+&\left(\mu\phi^{T}_{}i\tau_{2}^{}\Delta_{L}^{}\phi+\kappa\chi\eta^{T}_{}i\tau_{2}^{}\Delta_{L}^{}\eta+\textrm{h.c.}\right)\,,
\end{eqnarray}
where $\mu_{1,2,3}^{}$ and $\mu$ have the mass dimension-1, while
$\lambda_{1,...,5}^{}$, $\alpha_{1,2,3}^{}$, $\beta_{1,...,6}^{}$
and $\kappa$ are all dimensionless, $M_{\Delta}^{2}$ is the positive
mass-square of the Higgs triplet. Without loss of generality, $\mu$
and $\kappa$ will be conveniently set as real after proper phase
rotations.

\textit{The vacuum expectation values}: For $\lambda_{1}^{}>0$ and
$\mu_{1}^{2}<0$, we can guarantee that before the electroweak phase
transition, the real scalar $\chi$ acquires a nonzero vacuum
expectation value (VEV),
\begin{eqnarray}
\langle\chi\rangle\equiv u=
\sqrt{-\frac{\mu_{1}^{2}}{\lambda_{1}^{}}}\,.
\end{eqnarray}
We can then write the field $\chi$ in terms of the real physical
field $\sigma$ as
\begin{eqnarray}
\chi\equiv \sigma + u\,,
\end{eqnarray}
so that the explicit form of the Yukawa couplings become
\begin{eqnarray}
\label{lagrangian2} -\mathcal{L}&\supset&
y_{ij}^{}\overline{\psi_{Li}^{}}\eta
\nu_{Rj}^{}+\frac{1}{2}M_{ij}^{}\overline{\nu^{c}_{Ri}}\nu_{Rj}^{}+\frac{1}{2}f_{ij}^{}\overline{\psi_{Li}^{c}}i\tau_{2}^{}\Delta_{L}^{}\psi_{Lj}^{}\nonumber\\
&+&\mu\phi^{T}_{}i\tau_{2}^{}\Delta_{L}^{}\phi+\tilde{\mu}\eta^{T}_{}i\tau_{2}^{}\Delta_{L}^{}\eta+\frac{1}{2}g_{ij}^{}\sigma\overline{\nu_{Ri}^{c}}\nu_{Rj}^{}\nonumber\\
&+&\kappa\sigma\eta^{T}_{}i\tau_{2}^{}\Delta_{L}^{}\eta+\textrm{h.c.}+M^{2}_{\Delta}\textrm{Tr}\left(\Delta_{L}^{\dagger}\Delta_{L}^{}\right)\,,
\end{eqnarray}
where we defined,
\begin{eqnarray}
M_{ij}^{}\equiv g_{ij}^{}u \quad \textrm{and} \quad\tilde{\mu}\equiv
\kappa u .
\end{eqnarray}
For convenience, we diagonalize
$g_{ij}^{}\rightarrow g_{i}^{}$ as well as $M_{ij}^{}\rightarrow
M_{i}^{}$ by redefining $\nu_{Ri}^{}$ and then simplify the
Lagrangian (\ref{lagrangian2}) as
\begin{eqnarray}
\label{lagrangian3} -\mathcal{L}&\supset&
y_{ij}^{}\overline{\psi_{Li}^{}}\eta
N_{j}^{}+\frac{1}{2}f_{ij}^{}\overline{\psi_{Li}^{c}}i\tau_{2}^{}\Delta_{L}^{}\psi_{Lj}^{}+\mu\phi^{T}_{}i\tau_{2}^{}\Delta_{L}^{}\phi\nonumber\\
&+&\tilde{\mu}\eta^{T}_{}i\tau_{2}^{}\Delta_{L}^{}\eta+\kappa\sigma\eta^{T}_{}i\tau_{2}^{}\Delta_{L}^{}\eta+\textrm{h.c.}\nonumber\\
&+&\frac{1}{2}g_{i}^{}\sigma\overline{N_{i}^{}}N_{i}^{}+\frac{1}{2}M_{i}^{}\overline{N_{i}^{}}N_{i}^{}+M^{2}_{\Delta}\textrm{Tr}\left(\Delta_{L}^{\dagger}\Delta_{L}^{}\right)
\end{eqnarray}
with
\begin{eqnarray}
N_{i}^{}\equiv\nu_{Ri}^{}+\nu_{Ri}^{c}
\end{eqnarray}
being the heavy Majorana neutrinos.

After the electroweak symmetry breaking, we denote the different
VEVs as $\langle
\phi\rangle \equiv \frac{1}{\sqrt{2}}v$, $\langle \eta\rangle \equiv
\frac{1}{\sqrt{2}}v'$, $\langle \Delta_{L}^{}\rangle \equiv
\frac{1}{\sqrt{2}}v_{L}^{}$ and $\langle \chi\rangle \equiv u'$ and
then analyze the potential as a function of these VEVs,
\begin{eqnarray}
\label{potential2}
&&V(u',v,v',v_{L}^{})\nonumber\\
&=&\frac{1}{2}\mu_{1}^{2}u'^{2}_{}+\frac{1}{4}\lambda_{1}^{}u'^{4}_{}+\frac{1}{2}\mu_{2}^{2}v^{2}_{}+\frac{1}{4}\lambda_{2}^{}v^{4}_{}\nonumber\\
&+&\frac{1}{2}\mu_{3}^{2}v'^{2}_{}+\frac{1}{4}\lambda_{3}^{}v'^{2}_{}+\frac{1}{2}M^{2}_{\Delta}v_{L}^{2}+\frac{1}{4}(\lambda_{4}^{}+\lambda_{5}^{})v_{L}^{4}\nonumber\\
&+&\frac{1}{2}\alpha_{1}^{}u'^{2}_{}v^{2}_{}+\frac{1}{2}\alpha_{2}^{}u'^{2}_{}v'^{2}_{}+\frac{1}{2}\alpha_{3}^{}u'^{2}_{}v_{L}^{2}\nonumber\\
&+&\frac{1}{2}\left(\beta_{1}^{}+\beta_{2}^{}\right)v^{2}_{}v'^{2}_{}+\frac{1}{2}\left(\beta_{3}^{}+\beta_{4}^{}\right)v^{2}_{}v^{2}_{L}\nonumber\\
&+&\frac{1}{2}\left(\beta_{5}^{}+\beta_{6}^{}\right)v'^{2}_{}v^{2}_{L}+\frac{1}{\sqrt{2}}\mu
v^{2}_{}v^{}_{L}+\frac{1}{\sqrt{2}}\tilde{\mu}'v'^{2}_{}v_{L}^{}
\end{eqnarray}
with $\tilde{\mu}'\equiv\kappa u'$. Using the extremum conditions,
$0=\partial V /\partial u' =
\partial V/\partial v =\partial V/ \partial v' =\partial V /\partial
v_{L}^{}$, we obtain,
\begin{eqnarray}
\label{potential3}
0&=&\lambda_{1}^{}u'^{3}_{}+\mu_{1}^{2}u'+\alpha_{1}^{}
v^{2}_{}u'+\alpha_{2}^{} v'^{2}_{}u'+\alpha_{3}^{}
v_{L}^{2}u'\nonumber\\
&+&\frac{1}{\sqrt{2}}\kappa v'^{2}_{}v_{L}^{}\,,\\
0&=&\mu_{2}^{2}+\alpha_{1}^{}
u'^{2}_{}+\left(\beta_{1}^{}+\beta_{2}^{}\right)
v'^{2}_{}+\left(\beta_{3}^{}+\beta_{4}^{}\right)v_{L}^{2}\nonumber\\
&+&2\sqrt{2}\mu
v_{L}^{}+\lambda_{2}^{}v^{2}_{}\,,\\
0&=&\mu_{3}^{2}+\alpha_{2}^{}
u'^{2}_{}+\left(\beta_{1}^{}+\beta_{2}^{}\right)
v^{2}_{}+\left(\beta_{5}^{}+\beta_{6}^{}\right)v_{L}^{2}\nonumber\\
&+&2\sqrt{2}\tilde{\mu}'v_{L}^{}+\lambda_{3}^{}v'^{2}_{}\,,\\
0&=& \frac{1}{\sqrt{2}}\mu v^{2}_{}+\frac{1}{\sqrt{2}}\tilde{\mu}'
v'^{2}_{}+\left[
M_{\Delta}^{2}+\alpha_{3}u'^{2}_{}+\left(\beta_{3}^{}+\beta_{4}^{}\right)v^{2}_{}\right.\nonumber\\
&+&\left.\left(\beta_{5}^{}+\beta_{6}^{}\right)v'^{2}_{}\right]v^{}_{L}+\left(\lambda_{4}^{}+\lambda_{5}^{}\right)v_{L}^{3}\,.
\end{eqnarray}
For
\begin{eqnarray}
&&\left\{
\begin{array}{l}
\lambda_{3}^{}>0\,,\vspace*{2mm} \\
\mu_{3}^{2}+\alpha_{2}^{}
u'^{2}_{}+\left(\beta_{1}^{}+\beta_{2}^{}\right)
v^{2}_{}+\left(\beta_{5}^{}+\beta_{6}^{}\right)v_{L}^{2}\\
+2\sqrt{2}\tilde{\mu}'v_{L}^{}>0\,,
\end{array} \right.
\end{eqnarray}
the new Higgs doublet $\eta$ gets a zero VEV, i.e., $v'=0$. We
assume $\mu<M_{\Delta}^{}$ and $v^{2}_{}\ll
M_{\Delta}^{2},\,u'^{2}_{}$, and then deduce
\begin{eqnarray}
\label{vev2} v_{L}^{}&\simeq& \frac{1}{\sqrt{2}}\frac{\mu
v^{2}_{}}{M_{\Delta}^{2}+\alpha_{3}u'^{2}_{}+\left(\beta_{3}^{}+\beta_{4}^{}\right)v^{2}_{}}\nonumber\\
&\simeq& \frac{1}{\sqrt{2}}\frac{\mu
v^{2}_{}}{M_{\Delta}^{2}+\alpha_{3}u'^{2}_{}}\nonumber\\
&\simeq&\frac{1}{\sqrt{2}}\frac{\mu v^{2}_{}}{M_{\Delta}^{2}}\quad
\textrm{for} \quad M_{\Delta}^{2}\gg \alpha_{3}u'^{2}_{}\,.
\end{eqnarray}
Subsequently, $u'$ and $v$ can be solved,
\begin{eqnarray}
u'&=&\sqrt{-\frac{\mu_{1}^{2}+\alpha_{1}^{} v^{2}_{}+\alpha_{3}^{}
v_{L}^{2}}{\lambda_{1}^{}}}\nonumber\\
&\simeq&\sqrt{-\frac{\mu_{1}^{2}+\alpha_{1}^{}
v^{2}_{}}{\lambda_{1}^{}}}\,,\\
v&=&\sqrt{-\frac{\mu_{2}^{2}+\alpha_{1}^{}
u^{'2}_{}+\left(\beta_{3}^{}+\beta_{4}^{}\right)v_{L}^{2}+2\sqrt{2}\mu
v_{L}^{}}{\lambda_{2}^{}}}\nonumber\\
&\simeq&\sqrt{-\frac{\mu_{2}^{2}+\alpha_{1}^{}
u^{'2}_{}}{\lambda_{2}^{}}}\,,
\end{eqnarray}
for
\begin{eqnarray}
&&\left\{
\begin{array}{l}
\lambda_{1}^{}>0\,,\\
\mu_{1}^{2}+\alpha_{1}^{} v^{2}_{}+\alpha_{3}^{} v_{L}^{2}<0\,,
\end{array} \right.\\
\nonumber\\
&&\left\{
\begin{array}{l}
\lambda_{2}^{}>0\,,\\
\mu_{2}^{2}+\alpha_{1}^{}
u^{'2}_{}+\left(\beta_{3}^{}+\beta_{4}^{}\right)v_{L}^{2}+2\sqrt{2}\mu
v_{L}^{}<0\,,
\end{array} \right.
\end{eqnarray}

We then obtain the masses of resulting physical scalar bosons after
the electroweak symmetry breaking,
\begin{eqnarray}
M^{2}_{\delta^{++}_{}}&\simeq&M^{2}_{\Delta}+\alpha_{3}^{}u'^{2}_{}+\left(\beta_{3}^{}+\beta_{4}^{}\right)v^{2}_{}\,,\\
M^{2}_{\delta^{+}_{}}&\simeq&M^{2}_{\Delta}+\alpha_{3}^{}u'^{2}_{}+\left(\beta_{3}^{}+\frac{1}{2}\beta_{4}^{}\right)v^{2}_{}\,,\\
M^{2}_{\delta^{0}_{}}&\simeq&M^{2}_{\Delta}+\alpha_{3}^{}u'^{2}_{}+\beta_{3}^{}v\,,\\
m^{2}_{\eta_{}^{\pm}}&\simeq&\mu_{3}^{2}+\alpha_{2}^{}u'^{2}_{}+\beta_{1}^{}v^{2}_{}\,,\\
m^{2}_{\eta_{R}^{0}}&\simeq&\overline{m}_{\eta^{}_{}}^{2}+\delta{m}_{\eta^{}_{}}^{2}\,,\\
m^{2}_{\eta_{I}^{0}}&\simeq&\overline{m}_{\eta^{}_{}}^{2}-\delta{m}_{\eta^{}_{}}^{2}\,,\\
\label{massh1}
m^{2}_{h_{1}^{}}&\simeq& \overline{m}^{2}_{h}-\delta m^{2}_{h}\,,\\
\label{massh2} m^{2}_{h_{2}^{}}&\simeq& \overline{m}^{2}_{h}+\delta
m^{2}_{h}\,,
\end{eqnarray}
with
\begin{eqnarray}
\label{masseta1} \overline{m}_{\eta^{}_{}}^{2}&\equiv&
\mu_{3}^{2}+\alpha_{2}^{}u'^{2}_{}+\left(\beta_{1}^{}+\beta_{2}^{}\right)v^{2}_{}\,,\\
\label{masseta2} \delta{m}_{\eta^{}_{}}^{2}&\equiv&
\frac{\tilde{\mu}'\mu}{M_{\Delta}^{2}+\alpha_{3}^{}u'^{2}_{}}v^{2}_{}\simeq \frac{\tilde{\mu}'\mu}{M_{\Delta}^{2}}v^{2}_{}\,,\\
\overline{m}^{2}_{h}&\equiv&\lambda_{1}^{}u'^{2}_{}+\lambda_{2}^{}v^{2}_{}\,,\\
\delta
m^{2}_{h}&\equiv&\left[\left(\lambda_{1}^{}u'^{2}_{}-\lambda_{2}^{}v^{2}_{}\right)^{2}_{}
+4\alpha_{1}^{2}u'^{2}_{}v^{2}_{}\right]^{\frac{1}{2}}_{}\,.
\end{eqnarray}
Here $\eta^{+}_{}$ and $\eta_{R,I}^{0}$ are defined by
\begin{eqnarray}
\eta^{+}_{}&\equiv&\left(\eta^{-}_{}\right)^{\ast}_{}\,,\\
\eta^{0}_{}&\equiv&\frac{1}{\sqrt{2}}\left(\eta^{0}_{R}+i\eta^{0}_{I}\right)\,.
\end{eqnarray}
In addition, the mass eigenstates $h_{1,2}^{}$ are the linear
combinations of $h$ and $\sigma'$, i.e.,
\begin{eqnarray}
h_{1}^{}&\equiv& \sigma'\sin\vartheta+h\cos\vartheta\,,\\
h_{2}^{}&\equiv& \sigma'\cos\vartheta-h\sin\vartheta\,,
\end{eqnarray}
where $h$, $\sigma'$ are defined by
\begin{eqnarray}
\phi&\equiv&\frac{1}{\sqrt{2}}\left\lgroup
\begin{array}{c}
v +h\\
0\end{array} \right\rgroup\,,\quad \chi \equiv u'+\sigma'\,,
\end{eqnarray}
and the mixing angle is given by
\begin{eqnarray}
\label{angle} \tan 2\vartheta
\simeq\frac{2\alpha_{1}^{}u'v}{\lambda_{2}^{}v^{2}_{}-\lambda_{1}^{}u'^{2}_{}}\,.
\end{eqnarray}

\begin{figure}
\vspace{3cm} \epsfig{file=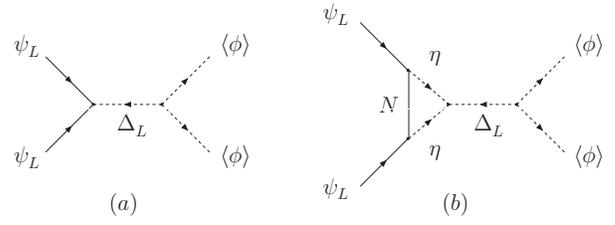, bbllx=4cm, bblly=6.0cm,
bburx=14cm, bbury=16cm, width=6cm, height=6cm, angle=0, clip=0}
\vspace{-6cm} \caption{\label{massgeneration} The neutrino
mass-generation. (a) is the type-II seesaw. (b) is the radiative
contribution.}
\end{figure}

\textit{Neutrino masses}: The first diagram of Fig.
\ref{massgeneration} shows the type-II seesaw approach to the
generation of the neutrino masses. It is reasonable to take the
scalar cubic coupling $\mu$ less than the triplet mass
$M_{\Delta}^{}$ in (\ref{vev2}). In consequence, the triplet VEV in
(\ref{vev2}) is seesaw-suppressed by the ratio of the electroweak
scale $v$ over the heavy mass $M_{\Delta}^{}$. Substantially, the
neutrinos naturally obtain the small Majorana masses,
\begin{eqnarray}
\label{mass2} (m_{\nu}^{II})_{ij}^{}\equiv
\frac{1}{\sqrt{2}}f_{ij}^{}v_{L}^{}\simeq -f_{ij}^{}\frac{\mu
v^{2}_{}}{ 2M^{2}_{\Delta}}\,.
\end{eqnarray}

For the zero VEV of new Higgs doublet $\eta$, we can not realize the
neutrino masses via the type-I seesaw. However, similar to
\cite{ma2006}, it is possible to generate the radiative neutrino
masses at one-loop order due to the trilinear scalar interactions in
(\ref{lagrangian3}). As shown in the second diagram of Fig.
\ref{massgeneration}, the one-loop process will induce a
contribution to the neutrino masses,
\begin{eqnarray}
\label{mass1}
(\widetilde{m}_{\nu}^{I})_{ij}^{}&=&\frac{1}{16\pi^{2}_{}}\sum_{k}^{}y_{ik}^{}y_{jk}^{}M'^{}_{k}\left[\frac{m_{\eta_{R}^{0}}^{2}}{m_{\eta_{R}^{0}}^{2}-M'^{2}_{k}}\ln\left(\frac{m_{\eta_{R}^{0}}^{2}}{M'^{2}_{k}}\right)\right.\nonumber\\
&-&\left.\frac{m_{\eta_{I}^{0}}^{2}}{m_{\eta_{I}^{0}}^{2}-M'^{2}_{k}}\ln\left(\frac{m_{\eta_{I}^{0}}^{2}}{M'^{2}_{k}}\right)\right]\,.
\end{eqnarray}
Here $M'^{}_{k}\equiv \frac{u'}{u}M^{}_{k}$. For $|\mu_{1}^{2}|\gg
|\alpha_{1}|v^{2}_{}$, we have $u'\simeq u$ and then
$M'^{}_{k}\simeq M^{}_{k}$, so the above formula can be simplified
as
\begin{eqnarray}
\label{mass11}
(\widetilde{m}_{\nu}^{I})_{ij}^{}&\simeq&\frac{1}{16\pi^{2}_{}}\sum_{k}^{}y_{ik}^{}y_{jk}^{}\frac{1}{M_{k}^{}}\nonumber\\
&\times&\left[m_{\eta_{R}^{0}}^{2}\ln\left(\frac{M_{k}^{2}}{m_{\eta_{R}^{0}}^{2}}\right)\right.-\left.m_{\eta_{I}^{0}}^{2}\ln\left(\frac{M_{k}^{2}}{m_{\eta_{I}^{0}}^{2}}\right)\right]\,.
\end{eqnarray}
by taking $m_{\eta^{0}_{R,I}}^{2} \ll M_{k}^{2}$. Moreover, from
(\ref{masseta1}) and (\ref{masseta2}), if $|\tilde{\mu}'\mu|\ll
M_{\Delta}^{2}$, we have $\delta m_{\eta^{}_{}}^{2} \ll
\overline{m}_{\eta^{}_{}}^{2}$ and then obtain
\begin{eqnarray}
\label{mass12}
(\widetilde{m}_{\nu}^{I})_{ij}^{}&\simeq&-\frac{1}{8\pi^{2}_{}}\sum_{k}^{}y_{ik}^{}y_{jk}^{}\frac{\delta
m_{\eta^{}_{}}^{2}}{M_{k}^{}}\left[1-\ln\left(\frac{M_{k}^{2}}{\overline{m}_{\eta^{}_{}}^{2}}\right)\right]\nonumber\\
&=&-\xi\sum_{k}^{}y_{ik}^{}y_{jk}^{}\frac{v^{2}_{}}{2M_{k}^{}}
\end{eqnarray}
for
\begin{eqnarray}
\label{xi} \xi&=&\mathcal{O}\left(\frac{1}{4\pi^{2}_{}}\frac{\delta
m_{\eta^{}_{}}^{2}}{v^{2}_{}}\left[1-\ln\left(\frac{M_{k}^{2}}{\overline{m}_{\eta^{}_{}}^{2}}\right)\right]\right)
\nonumber\\
&=&\mathcal{O}\left(\frac{1}{4\pi^{2}_{}}\frac{\tilde{\mu}'\mu}{M^{2}_{\Delta}}\left[1-\ln\left(\frac{M_{k}^{2}}{\overline{m}_{\eta^{}_{}}^{2}}\right)\right]\right)\,.
\end{eqnarray}
Note that the above loop-contribution will be absent once the values
of $\kappa$ and then $\tilde{\mu}'$ are taken to be zero.

\textit{Baryon asymmetry}: We now demonstrate how the observed
baryon asymmetry is generated in this model.
In the Lagrangian (\ref{lagrangian3}),
the lepton number of the left-handed lepton doublets and the Higgs
triplet are $1$ and $-2$, respectively, while those of the heavy
Majorana neutrinos, the Higgs doublets and the real scalar are all
zero. There are two sources of lepton number violation, one is the
trilinear interaction between the Higgs triplet and the Higgs
doublets, the other is the Yukawa couplings of the heavy Majorana
neutrinos to the left-handed lepton doublet and the new Higgs
doublet. Therefore, both the Higgs triplet and the heavy Majorana
neutrinos could decay to produce the lepton asymmetry if their
decays are CP-violation and out-of-equilibrium\footnote{Note that there
is an equivalent choice of lepton number:
$L=1$ for $\eta$ and $L=0$ for $\nu_R$, which makes only the $\mu \phi^T
i \tau_2 \Delta_L \phi$ term to be lepton number violating. So,
the CP asymmetry in the decays of $N_i$ and $\Delta_L$ can only
create an asymmetry in the numbers of $\psi_L$ and an equal and opposite
amount of asymmetry in the numbers of $\eta$. Thus there is no
net lepton number asymmetry
at this stage. However, since only the left-handed fields take part
in the sphaleron transitions, only the $\psi_L$ asymmetry
gets converted to a $B-L$ asymmetry before the electroweak phase
transition. After the electroweak phase transition, we are thus
left with a baryon asymmetry equivalent to the $B-L$ asymmetry
generated from the $\psi_L$ asymmetry and an equivalent amount
of $\eta$ asymmetry or lepton number asymmetry, which does not affect
the baryon asymmetry of the universe. In the rest of the article
we shall not discuss this possibility, since the final amount of
baryon asymmetry comes out to be the same.}.

\begin{figure}
\vspace{6.1cm} \epsfig{file=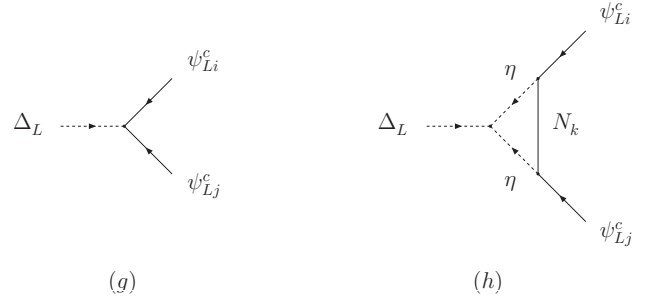, bbllx=4cm, bblly=6.0cm,
bburx=14cm, bbury=16cm, width=6cm, height=6cm, angle=0, clip=0}
\vspace{-7cm} \caption{\label{decay2} The Higgs triplets decay to
the leptons at one-loop order.}
\end{figure}

We can obtain the CP asymmetry in the decay of $N_{i}^{}$ through
the interference between the tree-level process and three one-loop
diagrams of Fig. \ref{decay2}, in which the first two one-loop
diagrams are the ordinary self-energy and vertex correction
involving another heavy Majorana neutrinos, while the third one-loop
diagram is mediated by the Higgs triplet \cite{odo1994}.
So it is convenient to
divide the total CP asymmetry into two independent parts,
\begin{eqnarray}
\label{CP2}
\varepsilon_{i}^{}&\equiv&\frac{\sum_{j}^{}\left[\Gamma\left(N_{i}^{}\rightarrow
\psi_{Lj}^{}\eta^{\ast}_{}\right)-\Gamma\left(N_{i}^{}\rightarrow
\psi_{Lj}^{c}\eta\right)\right]}{\Gamma_{i}}\nonumber\\
&=&\varepsilon_{i}^{N}+\varepsilon_{i}^{\Delta}\,,
\end{eqnarray}
where
\begin{eqnarray}
\label{decaywidth2}
\Gamma_{i}^{}&\equiv&\sum_{j}^{}\left[\Gamma\left(N_{i}^{}\rightarrow
\psi_{Lj}^{}\eta^{\ast}_{}\right)+\Gamma\left(N_{i}^{}\rightarrow
\psi_{Lj}^{c}\eta\right)\right]\nonumber\\
&=&\frac{1}{8\pi}\left(y^{\dagger}_{}y\right)_{ii}^{}M_{i}^{}
\end{eqnarray}
is the total decay width of $N_{i}^{}$, while
\begin{eqnarray}
\label{CP3}
\varepsilon_{i}^{N}&=&\frac{1}{8\pi}\frac{1}{\left(y^{\dagger}_{}
y\right)_{ii}}\sum_{k\neq i}^{}\textrm{Im}\left[\left(y^{\dagger}_{}
y\right)^{2}_{ik}\right]\nonumber\\
&\times&\sqrt{\frac{a_{k}^{}}{a_{i}^{}}}\left[1-\left(1+\frac{a_{k}^{}}{a_{i}^{}}\right)\ln\left(1+\frac{a_{i}^{}}{a_{k}^{}}\right)\right.\nonumber\\
&+&\left.\frac{a_{i}^{}}{a_{i}^{}-a_{k}^{}} \right]\,,\\
\label{CP4} \varepsilon_{i}^{\Delta}&=&\frac{3}{2
\pi}\frac{1}{\left(y^{\dagger}_{}y\right)_{ii}}\sum_{jm}^{}
\textrm{Im}\left( f^{\dagger}_{jm}
y^{\dagger}_{ij}y^{\dagger}_{im}\right)\frac{\tilde{\mu}}{M_{i}^{}}\nonumber\\
&\times&\left[1-\frac{a_{\Delta}^{}}{a_{i}^{}}\ln\left(1+\frac{a_{i}^{}}{a_{\Delta}^{}}\right)\right]
\end{eqnarray}
are the contributions of the first two one-loop diagrams and the
third one, respectively. Here the definitions
\begin{eqnarray}
a_{i}^{}\equiv \frac{M_{i}^{2}}{M_{1}^{2}}\,,\quad
a_{\Delta}^{}\equiv \frac{M_{\Delta}^{2}}{M_{1}^{2}}
\end{eqnarray}
have been adopted.

\begin{figure}
\vspace{6.1cm} \epsfig{file=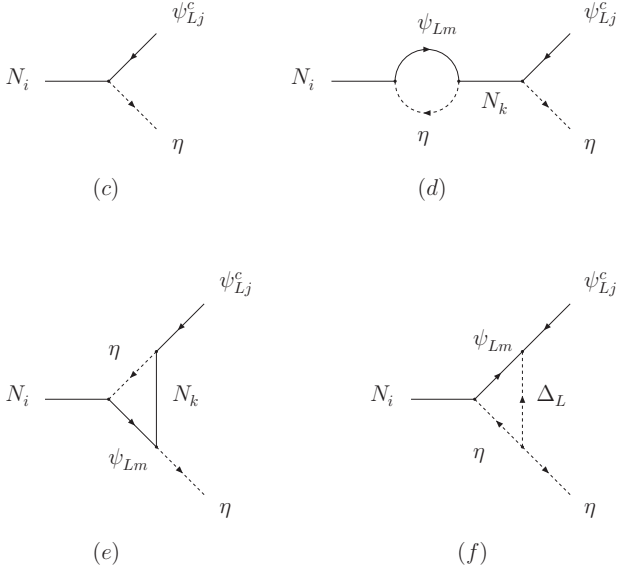, bbllx=4cm, bblly=6.0cm,
bburx=14cm, bbury=16cm, width=6cm, height=6cm, angle=0, clip=0}
\vspace{-3cm} \caption{\label{decay1} The heavy Majorana neutrinos
decay at one-loop order.  }
\end{figure}

Furthermore, as shown in Fig. \ref{decay1}, in the decay of
$\Delta_{L}^{}$, the tree-level diagram interferes with the one-loop
correction to generate the CP asymmetry,
\begin{eqnarray}
\label{cpv1} \varepsilon_{\Delta}^{}&\equiv&
2\frac{\sum_{ij}^{}\left[\Gamma\left(\Delta_{L}^{\ast}\rightarrow
\psi_{Li}^{}\psi_{Lj}^{}\right)-\Gamma\left(\Delta_{L}^{}\rightarrow
\psi_{Li}^{c}\psi_{Lj}^{c}\right)\right]}{\Gamma_{\Delta}^{}}\nonumber\\
&=&\frac{2}{\pi}\frac{\sum_{ijk}^{}\left(y^{}_{ki}y^{}_{kj}f_{ij}^{}\right)\tilde{\mu}M_{k}^{}
\ln\left(1+M_{\Delta}^{2}/M_{k}^{2}\right)}{\textrm{Tr}\left(f^{\dagger}_{}f\right)M_{\Delta}^{2}+4\tilde{\mu}^{2}_{}+4\mu^{2}_{}}
\end{eqnarray}
with
\begin{eqnarray}
\label{decaywidth1} \Gamma_{\Delta}^{}
&\equiv&\sum_{ij}^{}\Gamma\left(\Delta_{L}^{}\rightarrow
\psi_{Li}^{c}\psi_{Lj}^{c}\right)\nonumber\\
&+&\Gamma\left(\Delta_{L}^{}\rightarrow
\eta\eta\right)+\Gamma\left(\Delta_{L}^{}\rightarrow
\phi\phi\right)\nonumber\\
&\equiv&\sum_{ij}^{}\Gamma\left(\Delta_{L}^{\ast}\rightarrow
\psi_{Li}^{}\psi_{Lj}^{}\right)\nonumber\\
&+&\Gamma\left(\Delta_{L}^{\ast}\rightarrow
\eta_{}^{\ast}\eta_{}^{\ast}\right)+\Gamma\left(\Delta_{L}^{\ast}\rightarrow
\phi_{}^{\ast}\phi_{}^{\ast}\right)\nonumber\\
&=&\frac{1}{8\pi}\left[\frac{1}{4}\textrm{Tr}\left(f^{\dagger}_{}f\right)+\frac{\tilde{\mu}^{2}_{}+\mu^{2}_{}}{M_{\Delta}^{2}}\right]M_{\Delta}^{}
\end{eqnarray}
being the total decay width of $\Delta_{L}^{}$ or
$\Delta_{L}^{\ast}$.

Note that we have not considered the cases where $\sigma$ directly
decay to produce the leptons and anti-leptons through the imaginary
$N_{i}^{}$ or $\Delta_{L}^{}$ if $m_{\sigma}^{}>2M_{i}^{}$,
$M_{\Delta}^{}+2m_{\eta}^{}$ with $m_{\sigma}^{}$ and $m_{\eta}^{}$
being the masses of $\sigma$ and $\eta$, respectively. For
simplicity, here we will not discuss these cases.

It is straightforward to see that $\varepsilon_{\Delta}^{}$ and
$\varepsilon^{\Delta}_{i}$ will both be zero for $\kappa =0$ and
then $\tilde{\mu}=0$. In the following, to illustrate how to realize
non-resonant TeV leptogenesis, we first focus on the simple case
where $\varepsilon^{N}_{i}$ is the unique source of the CP
asymmetry. Note that $\tilde{\mu}'=0$ for $\kappa =0$, accordingly,
the one-loop diagram of Fig. \ref{massgeneration} is absent and
$N_{i}^{}$ have no possibility for the neutrino masses, we thus
obtain
\begin{eqnarray}
\label{CP5}
\varepsilon_{1}^{N}&\simeq&-\frac{3}{16\pi}\sum_{k=2,3}^{}\frac{\textrm{Im}\left[\left(y^{\dagger}_{}
y\right)^{2}_{1k}\right]}{\left(y^{\dagger}_{}
y\right)_{11}}\frac{M_{1}^{}}{M_{k}^{}}\nonumber\\
&\lesssim &-\frac{3}{16\pi}\left(\frac{M_{1}^{}}{M_{2}^{}}+\frac{M_{1}^{}}{M_{3}^{}}\right)\sin\delta\\
\end{eqnarray}
with $\delta$ being the CP phase. Here we have assumed $N_{1}^{}$ to be
the lightest heavy Majorana neutrinos, i.e., $M_{1}^{2}\ll
M_{2,3}^{2},\,M_{\Delta}^{2}$. The final baryon asymmetry can be
given by approximate relation \cite{kt1980}
\begin{eqnarray}
\label{asymmetry} Y_{B}^{}\equiv\frac{n_{B}^{}}{s}\simeq
-\frac{28}{79}\times\left\{
\begin{array}{ll}  \frac{\varepsilon_{1}^{}}{g_{\ast}^{}}\,,
&(\textrm{for}~K \ll
1)\,,\\
\frac{0.3\,\varepsilon_{1}^{}}{g_{\ast}^{}K\left(\ln
K\right)^{0.6}_{}}\,, &(\textrm{for}~K \gg 1)\,,
\end{array} \right.
\end{eqnarray}
where the factor $28/79$ is the value of $B/(B-L)$ and the parameter
$K$ is a measure of the departure from equilibrium and is defined by
\begin{eqnarray}
\label{parameter}
K&\equiv&\left.\frac{\Gamma_{1}^{}}{H(T)}\right|_{T=M_{1}^{}}=\left(y^{\dagger}_{}y\right)_{11}^{}\left(\frac{45}{2^{6}_{}\pi^{5}_{}g_{\ast}^{}}\right)^{\frac{1}{2}}_{}\frac{M_{\textrm{Pl}}^{}}{M_{1}^{}}\,.
\end{eqnarray}
Here
$H(T)=(4\pi^{3}_{}g_{\ast}^{}/45)^{\frac{1}{2}}T^{2}_{}/M_{\textrm{Pl}}^{}$
is the Hubble constant with the Planck mass $M_{\textrm{Pl}}^{}\sim
10^{19}_{}\,\textrm{GeV}$ and the relativistic degrees of freedom
$g_{\ast}^{}\sim 100$. For example, inspecting $M_{\Delta}^{}= 10\,
\textrm{TeV}$, $|\mu|=1\,\textrm{GeV}$ and $f\sim 10^{-6}_{}$ to
(\ref{mass1}), we obtain
$m_{\nu}^{}\sim\mathcal{O}(0.1\,\textrm{eV})$ which is consistent
with the neutrino oscillation experiments. Furthermore, let
$M_{1}=0.1\,M_{2,3}^{}=1\,\textrm{TeV}$, $y \sim 10^{-6}_{}$ and
$\sin\delta=10^{-3}_{}$, we drive the sample predictions: $K\simeq
48$ and $\varepsilon_{1}^{} \simeq - 1.2\times 10^{-5}_{}$. In
consequence,we arrive at $n_{B}^{}/s\simeq 10^{-10}$ as desired.

For $\kappa \neq 0$ and then $\tilde{\mu}', \tilde{\mu}\neq 0$,
$\Delta_{L}^{}$ and $N_{i}^{}$ will both contribute to the neutrino
masses and the lepton asymmetry. In the limit of $M_{\Delta}^{}\ll
M_{i}^{}$, the final lepton asymmetry is expected to mostly produce
by the decay of $\Delta_{L}^{}$. However, because the electroweak
gauge scattering should be out of thermal equilibrium, it is
difficult for a successful leptogenesis to lower the mass of
$\Delta_{L}^{}$ at TeV scale. Let us consider another possibility
that $N_{i}^{}$ are much lighter than $\Delta_{L}^{}$. In this case,
leptogenesis will be dominated by the decay of $N_{i}^{}$. For $
M_{1}^{2}\ll M_{2,3}^{2}\,,\,M_{\Delta}^{2}$ and
$|\tilde{\mu}'\mu|\ll M_{\Delta}^{2}$, $\varepsilon_{1}^{N}$ and
$\varepsilon_{1}^{\Delta}$ can be simplified as \cite{di2002}
\begin{eqnarray}
\label{CP5}
\varepsilon_{1}^{N}&\simeq&-\frac{3}{16\pi}\sum_{k=2,3}^{}\frac{\textrm{Im}\left[\left(y^{\dagger}_{}
y\right)^{2}_{1k}\right]}{\left(y^{\dagger}_{}
y\right)_{11}}\frac{M_{1}^{}}{M_{k}^{}}\nonumber\\
&\simeq& -\frac{3}{8 \pi}\frac{M_{1}^{}}{v^{2}_{}}\sum_{jk}^{}\frac{
\textrm{Im}\left[ (\widetilde{m}_{\nu}^{I\ast})_{jk}^{}
y^{\dagger}_{1j}y^{\dagger}_{1k}\right]}{\left(y^{\dagger}_{}y\right)_{11}}\frac{1}{\xi}\nonumber\\
&\simeq &-\frac{3}{8
\pi}\frac{M_{1}^{}\widetilde{m}_{\textrm{max}}^{I}}{v^{2}_{}}\frac{1}{\xi}\sin\delta'\,,\\
\label{CP6} \varepsilon_{1}^{\Delta}&\simeq&-\frac{3}{8
\pi}\frac{M_{1}^{}}{v^{2}_{}}\frac{\tilde{\mu}}{\mu}\sum_{jk}^{}\frac{
\textrm{Im}\left[ (m_{\nu}^{II\ast})_{jk}^{}
y^{\dagger}_{1j}y^{\dagger}_{1k}\right]}{\left(y^{\dagger}_{}y\right)_{11}}\nonumber\\
&\simeq &-\frac{3}{8
\pi}\frac{M_{1}^{}m_{\textrm{max}}^{II}}{v^{2}_{}}\left|\frac{\tilde{\mu}}{\mu}\right|\sin\delta''\,,
\end{eqnarray}
where $\delta'$ and $\delta''$ are CP phases,
$m_{\textrm{max}}^{II}$ and $\widetilde{m}_{\textrm{max}}^{I}$ are
the maximal eigenstates of the neutrino mass matrixes (\ref{mass2})
and (\ref{mass12}), respectively. Inputting $y\sim 10^{-7}_{}$,
$M_{1}=1\,\textrm{TeV}$ and $M_{2,3}^{}=10\,\textrm{TeV}$, we obtain
$\widetilde{m}_{\textrm{max}}^{I}=\mathcal{O}(10^{-3}_{}\,\textrm{eV})$.
Similarly, $m_{\textrm{max}}^{II}=\mathcal{O}(0.1\,\textrm{eV})$ for
$M_{\Delta}^{}=10\,\textrm{TeV}$, $|\mu|=1\,\textrm{GeV}$ and $f\sim
10^{-6}_{}$. Under this setup, we deduce $\xi\simeq 10^{-3}$ by
substituting $\overline{m}_{\eta}^{}=70\,\textrm{GeV}$,
$|\tilde{\mu}'|=10^{3}_{}\,\textrm{TeV}$ into (\ref{xi}) and then
have $\varepsilon_{1}^{N}\simeq -2\times 10^{-12}$ with the maximum
CP phase. We also acquire $\varepsilon_{1}^{\Delta}\simeq -3\times
10^{-8}$ for $|\tilde{\mu}|\simeq |\tilde{\mu}'|$ and
$\sin\delta''=0.15$. We thus drive the sample predictions: $K\simeq
0.5$ and $\varepsilon_{1}^{} \simeq \varepsilon_{1}^{\Delta}\simeq
-3\times 10^{-8}$. In consequence, we arrive at $n_{B}^{}/s\simeq
10^{-10}$ consistent with the cosmological observations.

\textit{Dark matter and Higgs phenomenology}: Since the new Higgs
doublet can not decay into the standard model particles, the neutral
$\eta^{0}_{R}$ and $\eta^{0}_{I}$ can provide the attractive
candidates for dark matter \cite{ma06,bhr2006,ma2006}. In particular, to
realize dark matter, $\eta^{0}_{R}$ and $\eta^{0}_{I}$ should have
the mass spectrum \cite{bhr2006}:
\begin{eqnarray}
\label{condition1}
\Delta m &\simeq &(8-\,\,\,9)\,\textrm{GeV}~~\textrm{for}~~ m_{L}=(60-73)\,\textrm{GeV}\,,\\
\label{condition2} \Delta m &\simeq& (9-
12)\,\textrm{GeV}~~\textrm{for}~~ m_{L}=(73- 75)\,\textrm{GeV}\,.
\end{eqnarray}
Here $\Delta m \equiv m_{NL}^{}-m_{L}^{}$ with $m_{L}^{}$ and
$m_{NL}^{}$ being the lightest and the next lightest masses between
$\eta^{0}_{R}$ and $\eta^{0}_{I}$. Note
\begin{eqnarray}
\overline{m}_{\eta}^{}&\equiv&
\frac{1}{2}\left(m_{L}^{}+m_{NL}^{}\right)\,, \\
|\delta
m_{\eta}^{2}|&\equiv&
\frac{1}{2}\left(m_{NL}^{2}-m_{L}^{2}\right)\,,
\end{eqnarray}
we thus deduce,
\begin{eqnarray}
m_{L}^{}&=&\overline{m}_{\eta}^{}\left(1-\frac{1}{2}\frac{|\delta m_{\eta}^{2}|}{\overline{m}_{\eta}^{2}}\right)\,,\\
\Delta m &=&\frac{|\delta m_{\eta}^{2}|}{\overline{m}_{\eta}^{}}\,.
\end{eqnarray}

In the previous discussions of TeV leptogenesis with $\kappa \neq
0$, we take $M_{\Delta}^{}=10\,\textrm{TeV}$,
$|\mu|=1\,\textrm{GeV}$, $|\tilde{\mu}|=10^{3}_{}\,\textrm{TeV}$ and
$\overline{m}_{\eta}^{}=70\,\textrm{GeV}$. It is straightforward to
see $|\delta m_{\eta}^{}| \simeq 25\, \textrm{GeV}$ from
(\ref{masseta2}). Therefore, we obtain $m_{L}\simeq 66\,
\textrm{GeV}$ and $\Delta m \simeq 9\,\textrm{GeV}$, which is
consistent with the mass spectrum (\ref{condition1}).

$\eta^{0}_{R}$ and $\eta^{0}_{I}$ are expected to be produced in
pairs by the standard model gauge bosons $W^{\pm}_{}$, $Z$ or
$\gamma$ and hence can be verified at the LHC. Once produced,
$\eta^{\pm}_{}$ will decay into $\eta_{R,I}^{0}$ and a virtual
$W^{\pm}_{}$, which becomes a quark-antiquark or lepton-antilepton
pair. For example, if $\eta^{0}_{R}$ is lighter than $\eta^{0}_{I}$,
the decay chain
\begin{eqnarray}
\eta^{+}_{}\rightarrow \eta^{0}_{I}l^{+}_{}\nu \,, \quad
\textrm{then} \quad \eta^{0}_{I}\rightarrow
\eta^{0}_{R}l^{+}_{}l^{-}_{}
\end{eqnarray}
has $3$ charged leptons and large missing energy, and can be
compared to the direct decay
\begin{eqnarray}
\eta^{+}_{}\rightarrow \eta^{0}_{R}l^{+}_{}\nu
\end{eqnarray}
to extract the masses of the respective particles.

As for the phenomenology of the Higgs triplet at the LHC as well as
the ILC, it has been discussed in \cite{mrs2000}. The same-sign
dileptons will be the most dominating modes of the $\delta^{++}_{}$.
Complementary measurements of $|f_{ij}^{}|$ at the ILC by the
process $e^{+}_{}e^{+}_{}(\mu^{+}_{}\mu^{-}_{})\rightarrow
l_{i}^{-}l_{j}^{-}$ would allow us to study the structure of the
neutrino mass matrix in detail.

For $\langle \chi\rangle=\mathcal{O}(\textrm{TeV})$, which is
natural to give the TeV Majorana masses of the right-handed
neutrinos and then realize the TeV leptogenesis, the mixing angle
$\vartheta$ and the splitting between $h_{1,2}^{}$ may be large.
Furthermore, the couplings of $h_{1,2}^{}$ to $W$ and $Z$ bosons,
quarks and charged leptons have essentially the same structure as
the corresponding Higgs couplings in the standard model, however,
their size is reduced by $\cos\vartheta$ and $\sin\vartheta$,
respectively. In the extreme case $\vartheta =\frac{\pi}{2}$, the
couplings of the lighter physical boson $h_{1}^{}$ to quarks and
leptons would even vanish. In other words, this mixing could lead to
significant impact on the Higgs searches at the LHC
\cite{bgm2006,bgc2006}.

\textit{Summary}: We propose a new model to realize leptogenesis and
dark matter at the TeV scale. A real scalar is introduced to
naturally realize the Majorana masses of the right-handed neutrinos.
Furthermore, we also consider a new Higgs doublet to provide the
attractive candidates for dark matter. Since the right-handed
neutrinos have no responsibility to generate the neutrino masses,
which is mostly dominated by the Higgs triplet through the type-II
seesaw, they can have large CP asymmetry at a low scale, such as
TeV, to produce the observed matter-antimatter asymmetry in the
universe, even if their Majorana masses are not highly
quasi-degenerate. It should be noticed that all new particles are
close to the TeV scale and hence should be observable at the LHC or
the ILC.


\begin{thebibliography}{99}

\bibitem{pdg2006}
Particle Data Group, W.M. Yao \textit{et al.}, Journal of Physics G
\textbf{33}, 1 (2006).


\bibitem{minkowski1977}
P. Minkowski, Phys. Lett. B \textbf{67}, 421 (1977); T. Yanagida, in
{\it Proc. of the Workshop on Unified Theory and the Baryon Number
of the Universe}, ed. O. Sawada and A. Sugamoto (KEK, Tsukuba,
1979), p. 95; M. Gell-Mann, P. Ramond, and R. Slansky, in {\it
Supergravity}, ed. F. van Nieuwenhuizen and D. Freedman (North
Holland, Amsterdam, 1979), p. 315; S.L. Glashow, in {\it Quarks and
Leptons}, ed. M. L$\rm\acute{e}$vy {\it et al.} (Plenum, New York,
1980), p. 707; R.N. Mohapatra and G. Senjanovi$\rm\acute{c}$, Phys.
Rev. Lett. \textbf{44}, 912 (1980); J. Schechter and J.W.F. Valle,
Phys. Rev. D \textbf{22}, 2227 (1980).

\bibitem{mp2002}
H. Murayama and A. Pierce, Phys. Rev. Lett. \textbf{89}, 271601
(2002); B. Thomas and M. Toharia, Phys. Rev. D \textbf{73}, 063512
(2006); B. Thomas and M. Toharia, Phys. Rev. D \textbf{75}, 013013
(2007); D.G. Cerdeno, A. Dedes, and T.E.J. Underwood, JHEP
\textbf{0609}, 067 (2006); P.H. Gu and H.J. He, JCAP \textbf{0612},
010 (2006).


\bibitem{fy1986}
M. Fukugita and T. Yanagida, Phys. Lett. B \textbf{174}, 45 (1986);
P. Langacker, R.D. Peccei, and T. Yanagida, Mod. Phys. Lett. A
\textbf{1}, 541 (1986).

\bibitem{luty1992}
M.A. Luty, Phys. Rev. D \textbf{45}, 455
(1992); R.N. Mohapatra and X. Zhang, Phys. Rev. D \textbf{46}, 5331
(1992).

\bibitem{fps1995}
M. Flanz, E.A. Paschos, and U. Sarkar, Phys. Lett. B
\textbf{345}, 248 (1995); M. Flanz, E.A. Paschos, U. Sarkar, and J.
Weiss, Phys. Lett. B \textbf{389}, 693 (1996).

\bibitem{pilaftsis1997}
A. Pilaftsis, Phys. Rev. D \textbf{56}, 5431 (1997); A. Pilaftsis
and T.E.J. Underwood, Nucl. Phys. B. \textbf{692}, 303 (2004).

\bibitem{ms1998}
E. Ma and U. Sarkar, Phys. Rev. Lett. \textbf{80}, 5716 (1998).



\bibitem{ars1998}
E.Kh. Akhmedov, V.A. Rubakov, and A.Yu. Smirnov, Phys. Rev. Lett.
\textbf{81}, 1359 (1998); K. Dick, M. Lindner, M. Ratz, and D.
Wright, Phys. Rev. Lett. \textbf{84}, 4039 (2000).


\bibitem{di2002}
S. Davidson and A. Ibarra, Phys. Lett. B \textbf{535}, 25 (2002); W.
Buchm$\rm\ddot{u}$ller, P. Di Bari, and M. Pl$\rm\ddot{u}$macher,
Nucl. Phys. B \textbf{665}, 445 (2003); Hambye and G.
Senjanovi$\rm\acute{c}$, Phys. Lett. B \textbf{582}, 73 (2004); S.
Antusch, S.F. King, Phys. Lett. B \textbf{597}, 199 (2004); P. Gu
and X.J. Bi, Phys. Rev. D \textbf{70}, 063511 (2004).

\bibitem{guo2004}
W.L. Guo, Phys. Rev. D \textbf{70}, 053009 (2004); G. D'Ambrosio
\textit{et al.}, Phys. Lett. B \textbf{604}, 199 (2004); S. Antusch,
S.F. King, JHEP \textbf{0601}, 117 (2006); N. Sahu, U. Sarkar, Phys.
Rev. D \textbf{74}, 093002 (2006).

\bibitem{bpy2005}
W. Buchmuller, R.D. Peccei, and T. Yanagida, Ann. Rev. Nucl. Part.
Sci. \textbf{55}, 311 (2005).




\bibitem{ma06} E. Ma, Phys. Rev. D {\bf 73}, 077301 (2006).

\bibitem{bhr2006}
R. Barbieri, L.J. Hall, and V.S. Rychkov, Phys. Rev. D \textbf{74},
015007 (2006).




\bibitem{ma2006}
E. Ma, Mod. Phys. Lett. A \textbf{21}, 1777 (2006).


\bibitem{co2006}
X. Calmet and J.F. Oliver, Europhys. Lett. \textbf{77}, 51002
(2007).

\bibitem{ma77} E. Ma, S. Pakvasa, and S.F. Tuan, Phys. Rev. D {\bf 16}, 1568
(1977); N.G. Deshpande and E. Ma, Phys. Rev. D {\bf 18}, 2574
(1978).

\bibitem{hnot2006}
L.L. Honorez, E. Nezri, J.F. Oliver, and M.H.G. Tytgat, JCAP
\textbf{0702}, 028 (2007); M. Gustafsson, E. Lundstrom, L.
Bergstrom, and J. Edsjo, arXiv:astro-ph/0703512.

\bibitem{kuzmin1997}
V.A. Kuzmin, arXiv:hep-ph/9701269.


\bibitem{krs1985}
V.A. Kuzmin, V.A. Rubakov, and M.E. Shaposhnikov, Phys. Lett. B
\textbf{155}, 36 (1985); R.N. Mohapatra and X. Zhang, Phys. Rev. D
\textbf{45}, 2699 (1992).

\bibitem{odo1994} P.J. O'Donnell and U. Sarkar, Phys. Rev. D
{\bf  49}, 2118 (1994).



\bibitem{kt1980}
E.W. Kolb and M.S. Turner, \textit{The Early Universe},
Addison-Wesley, Reading, MA, 1990; H.B. Nielsen and Y. Takanishi,
Phys. Lett. B \textbf{507}, 241 (2001).



\bibitem{mrs2000}
E. Ma, M. Raidal, and U. Sarkar, Phys. Rev. Lett. \textbf{85}, 3769
(2000); Nucl. Phys. B \textbf{615}, 313 (2001).


\bibitem{bgm2006}
W. Buchm$\rm\ddot{u}$ller, C. Greub, and P. Minkowski, Phys. Lett. B
\textbf{267}, 395 (1991).


\bibitem{bgc2006}
O. Bahat-Treidel, Y. Grossman, and Y.R. Comments,
arXiv:hep-ph/0611162; J.R. Espinosa and M. Quir$\rm\acute{o}$s,
arXiv:hep-ph/0701145.




\end{thebibliography}
\end{document}